\documentclass[final,letterpaper,twoside,11pt]{article}
\usepackage{graphicx}
\usepackage{subfigure}
\usepackage{bbm}
\def\lsim{\mathrel{\rlap{\lower4pt\hbox{\hskip1pt$\sim$}}
    \raise1pt\hbox{$<$}}}
\def\gsim{\mathrel{\rlap{\lower4pt\hbox{\hskip1pt$\sim$}}
    \raise1pt\hbox{$>$}}} 

\newcommand{\bra}[1]{ \langle {#1} | }
\newcommand{\ket}[1]{ | {#1} \rangle }

\newcommand{\kev}{{\rm keV}}
\newcommand{\mev}{{\rm MeV}}
\newcommand{\gev}{{\rm GeV}}
\newcommand{\tev}{{\rm TeV}}

\newcommand{\be}{\begin{eqnarray}}
\newcommand{\ee}{\end{eqnarray}}
\newcommand{\fslash}{\displaystyle{\not}}
\newcommand{\order}{\mathcal{O}}
\newcommand{\bolde}{\textbf{\scriptsize{e}}}
\newcommand{\boldp}{\textbf{\scriptsize{p}}}
\newcommand{\boldH}{\textbf{\scriptsize{H}}}

\begin{document}

\title{Atomic Dark Matter}
\author{David E. Kaplan\thanks{dkaplan@pha.jhu.edu}\, \ \,Gordan Z. Krnjaic\thanks{gordan@pha.jhu.edu}\, \ \,Keith R. Rehermann\thanks{keith@pha.jhu.edu} \ \,Christopher M. Wells\thanks{cwells13@pha.jhu.edu}\, \\ \\ \it Department of Physics and Astronomy\\ \it Johns Hopkins University\\ \it 3400 North Charles Street\\ \it Baltimore, MD 21218-2686}

\date{\today}
\maketitle
\begin{abstract}

We propose that dark matter is dominantly comprised of atomic bound states.  We build a simple
model and map the parameter space that results in the early universe
formation of hydrogen-like dark atoms.  We find that atomic dark matter has interesting
implications for cosmology as well as direct detection: Protohalo
formation can be suppressed below $M_{proto} \sim 10^3\, - \, 10^6 M_{\odot}$ for weak
scale dark matter due to Ion-Radiation interactions in the dark
sector. Moreover, weak-scale dark atoms can accommodate hyperfine splittings of order
$100\,\kev$, consistent with the inelastic dark matter interpretation of
the DAMA data while naturally evading direct detection bounds.
\end{abstract}

\section{Introduction}
Cosmological observations suggest that dark matter comprises more
than $80\%$ of the matter in the universe \cite{Komatsu:2008hk, Abazajian:2004tn}.  Much of
the effort to explain the origin of dark matter has focused on
minimal solutions in which dark matter consists of a single
particle species, the most popular being the neutralino in variants
of the supersymmetric standard model.  Such dark matter models include
the compelling feature that weak-scale physics --  weak-scale mass and
weak-force coupling strength -- can naturally generate dark matter with the
correct cosmological abundance.  Dark matter in this broad
class is described as weakly interacting massive  particles
(WIMPs). \\

However, conflicts do exist between WIMP models and
observational data.  Simulations of WIMP dark matter predict
significantly more small-scale structure than current observations
suggest \cite{Navarro:1995iw,Gilmore:2007fy}.  In addition, the direct detection
experiment, DAMA \cite{Bernabei:2008yi}, sees a positive signal with great
significance ($8\sigma$), yet when interpreted as a standard
WIMP, other experiments such as CDMS \cite{Ahmed:2008eu} and XENON10 
\cite{Angle:2007uj}, 
completely rule out the same parameter space.  Finally, measured
cosmic ray spectra may suggest a new primary source for electrons and
positrons in our galaxy and potentially evidence for dark-matter
annihilation; however, the standard neutralino candidate is unable to
fit this data \cite{Adriani:2008zr,:2008zzr,Torii:2008xu,Abdo:2009zk}.\\

These issues suggests compelling reasons to explore
dark matter models beyond the minimal candidate.  In addition, the
dark matter sector (or `dark sector') may be rich with complexity and
may feature unanticipated dynamics.  In fact, the dark
matter may even interact via a long-range force -- a massless gauge
boson -- which is still allowed by the bounds on the number of relativistic
degrees of freedom during big bang nucleosynthesis \cite{Amsler:2008zzb}.\\

In this paper we propose a dark
sector charged under a hidden U(1) gauge symmetry.  We assume two
species of fermions, a `dark proton' and a `dark electron', and that
the dark matter abundance comes from a matter--anti-matter
asymmetry.\footnote{ Some models that use the matter--anti-matter
  asymmetry to generate the correct dark matter abundance exist
  \cite{Kaplan:2009ag,Farrar:2005zd}, but we do
  not explore them here.
}
We shall see that in interesting parts of parameter space, the bulk
of the dark matter exists in atomic bound states.  The Lagrangian
is
\begin{equation}
\mathcal{L}_{dark} = \overline{\Psi}_{\textbf{\scriptsize{p}}}(\fslash D+m_{\textbf{\scriptsize{p}}})
  \Psi_{\textbf{\scriptsize{p}}} + \overline{\Psi}_{\textbf{\scriptsize{e}}}(\fslash
  D+m_{\textbf{\scriptsize{e}}})\Psi_{\textbf{\scriptsize{e}}}
\label{eq:Ldark}
\end{equation}
where $\fslash D=i\fslash \partial+g Q\fslash A$ and $Q = \pm 1$ for
$\Psi_{\textbf{\scriptsize{p}}}$ and $\Psi_{\textbf{\scriptsize{e}}}$ respectively.  In what follows we use the
convention $m_{\textbf{\scriptsize{p}}} \ge m_{\textbf{\scriptsize{e}}}$ without loss of generality.  We show
(Section 2) that for parts of parameter space, recombination in the
dark sector occurs efficiently, and we discuss the bounds from and
implications for structure formation.  We then add interactions
which allow for direct detection in a way that mimics inelastic dark
matter \cite{TuckerSmith:2001hy} and show that there exist parts of parameter space
which can explain the DAMA signal, while avoiding constraints from
other direct detection experiments (Section 3).  Finally, in Section 4
we discuss, in a cursory way, other phenomena potentially related to
atomic dark matter. \\ 

A number of ideas related to this work have
appeared in the literature.  For example, the idea of U(1) charged
dark matter has appeared in \cite{Feng:2009mn,Ackerman:2008gi}, the
idea of composite dark
matter in \cite{Alves:2009nf}, and that of mirror dark matter in
\cite{Berezhiani:2005ek, Mohapatra:2001sx}.  To our knowledge, this is
the first work to explore the generic parameter space for viable
atomic dark matter.

\section{Cosmology}
\label{sec:cosmo}
Introducing a new hidden U(1) has interesting cosmological
implications.  Our interests lie in the parameter space that affords
atomic systems.  The existence of standard model (SM) atomic hydrogen states in the early
universe requires an asymmetry between particles and
antiparticles; dark atoms are no different.  We assume that there is a
`dark asymmetry' akin to the baryon
asymmetry in the SM,
and that the dark asymmetry is such
that the universe is net charge neutral, $n_{\textbf{\scriptsize{e}}}=n_{\textbf{\scriptsize{p}}}$.\footnote{Unless
  otherwise noted, $\textbf{\scriptsize{e}}$,  $\textbf{\scriptsize{p}}$, and $\textbf{\scriptsize{H}}$ refer to the dark electron, dark
  proton, and dark hydrogen, respectively.} 
The existence of dark atoms implies that dark matter is
coupled to dark radiation until the universe cools
beyond the binding energy of hydrogen
\be
\label{eq:binding_energy}
B=\frac{1}{2}\alpha_D^2\mu_{\textbf{\scriptsize{H}}},
\ee
where $\alpha_D$ is the dark fine structure constant and
$\mu_{\textbf{\scriptsize{H}}}=(m_{\textbf{\scriptsize{e}}} m_{\textbf{\scriptsize{p}}})/(m_{\textbf{\scriptsize{e}}}+m_{\textbf{\scriptsize{p}}})$ is the reduced mass of dark hydrogen.
This has potentially interesting
implications for structure formation because interactions in the dark
sector can decouple much later than in a conventional CDM WIMP model.
Observations of satellite galaxies seem to favor some mechanism to
damp the growth of small scale structure in dark matter
\cite{Klypin:1999uc,Moore:1999nt}, which, as discussed below, can be
provided by atomic dark matter. \\

\subsection{Dark Recombination and Halo Constraints}
\label{sec:recomb}
One of the most interesting features of the model is the presence of
both neutral and ionized dark matter components.  
The fractional ionization, $X_{\textbf{\scriptsize{e}}}$, plays an important role in the cosmic
evolution of the dark matter.  At early
times, $X_{\textbf{\scriptsize{e}}}$ affects the decoupling temperature of dark matter and
dark radiation, which impacts small-scale structure formation of dark
matter.  At late times, bounds on 
dark matter self-interactions constrain $X_{\textbf{\scriptsize{e}}}$ because the
dark matter ions interact through a long range force.  
The residual ionization fraction in the dark sector is governed by
neutral atom formation in analogy with SM hydrogen recombination \cite{Peebles:1968ja}.
In the following, we follow the notation of Ref.\,\cite{Dodelson:2003}. \\

The residual ionization fraction is found by solving the
Boltzmann equation for the free dark electron fraction, 
\be
\label{eq:xe}
X_{\textbf{\scriptsize{e}}} \equiv \frac{n_{\textbf{\scriptsize{e}}}}{n_{\textbf{\scriptsize{e}}}+n_{\textbf{\scriptsize{H}}}}.
\ee
The evolution of $X_{\textbf{\scriptsize{e}}}$ depends on the Hubble rate, H, and the
rate for $\textbf{e}+\textbf{p} \leftrightarrow \textbf{H}+ \textbf{$\gamma$}$.
We can write the thermally-averaged recombination cross section
using
the dimensionless variable $x=\frac{B}{T}$ as
\begin{equation}
\label{eq:sigma_v}
\langle \sigma v \rangle
=\xi \frac{64 \pi}{\sqrt{27 \pi}} \frac{\alpha_D^2}{\mu_{\textbf{\scriptsize{H}}}^2}x^{1/2} {\rm ln}(x).
\end{equation}
where $\xi = 0.448 $ is a best-fit numerical coefficient
\cite{Ma:1995ey,Spitzer:1978}. The equation governing $X_{\textbf{\scriptsize{e}}}$ can be
written as  
\begin{equation}
\frac{dX_{\textbf{\scriptsize{e}}}}{dx} = 
        C\frac{1}{H x} 
	\left[(
              1-X_{\textbf{\scriptsize{e}}})^2 \beta
	      -X_{\textbf{\scriptsize{e}}}^2 n_{DM} \langle \sigma v \rangle 
	\right]
\label{eq:boltz_xe} 
\end{equation}\\
where
\be
\label{eq:beta}
	      \beta=
	      \langle \sigma v \rangle 
	      \left ( \frac{B m_{\textbf{\scriptsize{e}}}}{2\pi x} \right )^{3/2} {\rm e}^{-x}. 
\ee

\begin{figure}[t]
\begin{center}
\includegraphics[scale = 0.8]{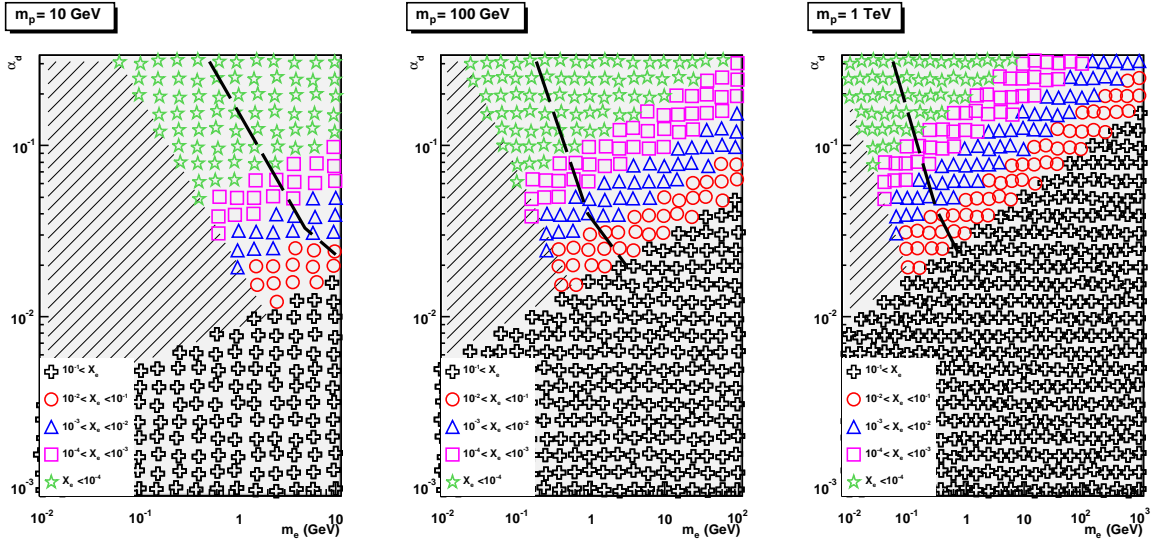}
\caption{The allowed parameter space in $\alpha_D-m_{\textbf{\scriptsize{e}}}$ for a given
  $m_{\textbf{\scriptsize{p}}}$ and as a function of the residual ionization fraction,
  $X_{\textbf{\scriptsize{e}}}$. Atom dark matter is viable in the colored regions, which
  correspond to $10^{-2}<X_{\textbf{\scriptsize{e}}}<10^{-1}$ (red circles), $10^{-3}<X_{\textbf{\scriptsize{e}}}<10^{-2}$
  (blue triangles), $\,10^{-4}<X_{\textbf{\scriptsize{e}}}<10^{-3}$
  (magenta boxes) and $X_{\textbf{\scriptsize{e}}}<10^{-4}$
  (green stars). The striped region is ruled out by
  Eq.\,\ref{eq:sigmaMbound} with $\kappa = 3$ and this region extends to the dashed black line for $\kappa = 10$. The black-crossed region is
  ruled out because $X_{\textbf{\scriptsize{e}}}> 10\%$.}
\label{fig:recomb}
\end{center}
\end{figure}

As discussed in \cite{Peebles:1968ja,Ma:1995ey}, recombination into the
$n=2$ state completely dominates the evolution of $X_{\textbf{\scriptsize{e}}}$. This is accounted for through the factor $C$ in Eq.\,(\ref{eq:boltz_xe})
which represents the fraction of $n=2$ states that produce a net gain
in the number of ground state hydrogen atoms.  This is not unity
because the thermal bath can ionize the $n=2$ state before it
decays. Thus, $C$ is the ratio of the $(n=2 \rightarrow n=1)$ decay
rate to the sum of this decay rate plus the ionization rate (see
\cite{Peebles:1968ja} for a detailed discussion)

\be
C= \frac{\Lambda_{\alpha}+\Lambda_{2\gamma}}{\Lambda_{\alpha}+\Lambda_{2\gamma}+\beta^{(2)}}.
\label{eq:C_n2}
\ee
The rates are given by
\be
\beta^{(2)}&=&\beta \rm{e}^{3x/4}\\
\Lambda_{\alpha}&=&H\frac{(3B)^3}{(8\pi)^2}\frac{1}{(1-X_{\textbf{\scriptsize{e}}})\,n_{DM}}\\
\Lambda_{2\gamma}&=&3.8\frac{3^2}{2^{11}}\frac{\alpha^8 \mu_{H}}{4 \pi}
\label{eq:rates_def}
\ee
where $\Lambda_{\alpha}$ the rate for a Lyman-$\alpha$ photon to
redshift such that it cannot excite $n=1 \rightarrow n=2$ and
$\Lambda_{2\gamma}$ is the two photon decay rate of $2s \rightarrow 1s$, which has been taken from
Ref. \cite{Spitzer:1951}. We find that $X_e$ varies from
$1-10^{-10}$ throughout the parameter space $\alpha_D \in
[10^{-3},0.3]$ , $m_{\bolde} \in [0.01 \, \gev, m_{\boldp}]$, $m_{\boldp} \in
[m_{\bolde}, 3 \, \tev]$. Self-interactions, as discussed below, rule out some of this
parameter space. A few representive planes are plotted in Figure \ref{fig:recomb}.\\

Bounds on the atomic parameter space and $X_e$ can be derived from
observations of the Bullet Cluster and halo profiles. The bounds are derived through the momentum transfer cross
section\footnote{Note that $\sigma_{mt}$ reduces to the total elastic
  cross section for hard-sphere s-wave scattering which is typical of WIMP dark matter models.}
\be
\sigma^{i}_{mt} = \int d\Omega(1-\cos\theta)\frac{d\sigma^{i}}{d\Omega}
\label{eq:sigma_mt}
\ee
where the index $i$ runs over the three types of self-interactions present in our dark sector: Hydrogen--Hydrogen, Ion--Hydrogen, and
Ion--Ion.  The last process is described by Coulomb scattering, but
since we want to study a dominantly atomic dark sector, the Ion--Ion
cross section is the least relevant to our model and we do not discuss
it further.  Na\"{i}vely, the first two cross sections are bounded by
geometric values 

\be
\frac{d\sigma}{d\Omega} \leq a_{0}^{2},
\ee
where $a_{0} = 1/(\alpha\,\mu_{\boldH})$ is the Bohr radius of dark hydrogen.
However, this na\"{i}ve guess is inadequate.  At low energies
($k\,a_{0} \ll 1$) both of these processes can be described by
scattering from a central potential 

\be
V \propto r^{-n}
\ee
with $n = 6\;\textrm{and}\;4$, respectively.  In this case one finds
that the cross sections are velocity-independent constants and enhanced over the geometric
estimate \cite{Landau:1958}.  We are generally interested in a wide
range of $ka_0$, thus these results are not strictly applicable
however these cross sections are slowly decreasing functions of the
relative velocity  \cite{Landau:1958}.  A conservative estimate of
the cross sections is given by

\be
\label{eq:enhancedsigma}
\frac{d\sigma}{d\Omega} \leq (\kappa\,a_{0})^{2},
\ee 
with $3\leq\,\kappa\,\leq 10$.  The values of $\kappa$ have been
inferred from general quantum mechanical scattering
\cite{Wu:1962,Bransden:1983} and detailed
computations of SM hydrogen scattering \cite{Schultz:2008apj,Krstic:1999jpb, Kristic:2004pra}.\\

With the relevant cross sections in hand, we can use observations of the Bullet Cluster \cite{Markevitch:2003at,Randall:2007ph} as a guide
for the present day maximum value of $X_{\textbf{\scriptsize{e}}}$.  Measurements of the mass-to-light ratio and
the radius of the sub-cluster suggest that the sub-cluster could have
lost no more than $F_{obs}=20-30 \%$ of its initial mass.  Following the
analysis in \cite{Markevitch:2003at}, the number of scattering centers
that a single dark matter particle encounters as it passes through
the  target cluster in the case of one species is 
\be
\label{eq:tau}
\tau=\frac{\sigma}{m} \Xi_s;
\ee
this quantity is often referred to as the scattering depth.
The parameter $\Xi_s$ is the surface mass density of the sub-cluster
defined as
\be
\Xi_{s} \equiv \int_{0}^{R}\rho(z)\,dz,
\ee
where $\rho(z)$ is the sub-cluster's volume mass density and $R$ is
the radius of the sub-cluster.  For
multiple species, with species $i$ in the sub-cluster having a mass
density $\Xi_i$, and mass $m_i$, scattering off of species $j$ in
the target cluster, we have 
\be
\label{eq:tau_ij}
\tau_{ij} = \left(\frac{\Xi}{m}\right)_i \sigma_{ij} f_j \;\;\;\;\; ({\rm no
  \, sum})
\ee
where $\sigma_{ij}$ is the cross section for $i$ and $j$ to interact and
$f_j$ is the number fraction of species $j$ in the target
cluster. Equation \ref{eq:tau_ij} can be rewritten in terms the of total
observed surface mass density, $\Xi_T = \displaystyle\sum_{i} \Xi_i$, as
\be
\label{eq:tau_tot}
\tau_{ij}=\Xi_T \frac{f_i}{\displaystyle\sum_{k} f_k m_k}\sigma_{ij}f_j.
\ee

We make the
simplifying and conservative assumption that all of the ions in the
sub-cluster are scattered out of the sub-cluster.  In this case, the
mass fraction lost through $H-j$ scattering is bounded by 
\be
\label{eq:mass_frac}
F_{\textbf{\scriptsize{H}}}=F_{obs}-f_{\textbf{\scriptsize{p}}}\frac{m_{\textbf{\scriptsize{p}}}}{m_{\textbf{\scriptsize{H}}}}-f_{\textbf{\scriptsize{e}}}\frac{m_{\textbf{\scriptsize{e}}}}{m_{\textbf{\scriptsize{H}}}}=F_{obs}-X_{\textbf{\scriptsize{e}}}.
\ee
The mass fraction actually lost from the sub-cluster, using cross
sections as parameterized in (\ref{eq:enhancedsigma}) and (\ref{eq:tau_tot}), is
\be
\label{eq:mass_lost}
\Delta = \displaystyle\sum_{j} \tau_{Hj} = \frac{\Xi_T}{m_{\textbf{\scriptsize{H}}}}
(1-X^2_{\textbf{\scriptsize{e}}}) 4\pi \kappa^{2}a_0^2.
\ee
Demanding that $\Delta < F_{\textbf{\scriptsize{H}}}$ we have the bound
\be
\label{eq:sigmaMbound}
\frac{\sigma}{m_{\textbf{\scriptsize{H}}}}=\kappa^2\frac{a_0^2}{m_{\textbf{\scriptsize{H}}}} < \frac{F_{obs}-X_{\textbf{\scriptsize{e}}}}{\Xi_T(1-X^2_{\textbf{\scriptsize{e}}})}.
\ee

Plugging in the values $\Xi_T = 0.2\,\textrm{--}\,0.3 \, \rm{cm}^{-2}\rm{g}$ and
$F_{obs}=0.2$ gives a constraint on the atomic parameter space \cite{Markevitch:2003at}
\be
\label{eq:sigmaMbound_numbers}
\left( \frac{0.1}{\alpha_D}\right)^2 
\left(\frac{1 \, \gev}{\mu_{\textbf{\scriptsize{H}}}}\right)^2  
\left(\frac{100\, \gev}{m_{\textbf{\scriptsize{H}}}} \right)  
\lsim  (20\,\textrm{--}\,200)\,\frac{0.2-X_{\textbf{\scriptsize{e}}}}{1-X^2_{\textbf{\scriptsize{e}}}}.
\ee

Thus, we find that $X_{\textbf{\scriptsize{e}}} $ and $\sigma/m_{\textbf{\scriptsize{H}}} $
are bounded simultaneously. From our
conservative (and representative) assumption about the Ion-Ion cross section,
$X_{\textbf{\scriptsize{e}}}$ is bounded to be less than $10\%-20\%$ regardless of
$\sigma/m_{\textbf{\scriptsize{H}}}$. For very small  $X_{\textbf{\scriptsize{e}}}$, the
usual CDM WIMP bounds on $\sigma/m_{\boldH}$ are applicable;
our estimate yields $\sigma/m_{\boldH} \lsim 1 \, \rm{cm^2/g}$, which
is slightly larger than the detailed simulations of \cite{Randall:2007ph}. 
Figure \ref{fig:recomb} shows some of the allowed parameter space in the
$\alpha_D-m_{\textbf{\scriptsize{e}}}$ plane for a few atomic masses ranging
between $10 \, \gev$ and $1 \, \tev$.\\

Previous considerations of a hidden $U(1)$
\cite{Feng:2009mn,Ackerman:2008gi} have concluded that soft scattering of charged
dark matter can drastically
affect halo formation and thereby rule out large swaths of parameter
space. This result follows from the
soft singularity in the Rutherford scattering rate which, when integrated over
galactic time scales, can lead to significant energy transfer between
charged particles.  This effect tends to smooth out the core of the dark matter distribution.  Application of these results excludes \emph{all} of the
parameter space shown in Figure\,\ref{fig:recomb}.  However, since Hydrogen-Hydrogen scattering is well modelled by
hard-sphere scattering in the majority of the considered parameter
space, these bounds are not applicable to atomic dark matter.  The relevant bounds from halo formation considerations are
$0.1\,\textrm{cm$^2$/g}\,\lsim\,\sigma/m_{\textbf{\scriptsize{e}}}\,\lsim\,1\,\textrm{cm$^2$/g}$,
which do not signficantly change our conclusions \cite{MiraldaEscude:2000qt}.  Atomic dark matter
provides a dynamical mechanism to shut off the na\"{i}ve long range effects of a
hidden U(1).
\subsection{Protohalos and Radiation Damping}

Certain values of the residual ionization fraction
may enable the ionized dark matter to smooth out structure on small
scales.  The parameter space of atomic dark matter allowed by
constrains derived in Section \ref{sec:recomb} 
predict that kinetic decoupling of the dark matter fluid and dark radiation occurs
during the radiation dominated epoch.  The dark radiation damps the
growth of structure until it decouples from the dark matter \cite{Loeb:2005pm}.  The
decoupling temperature is given by
\be
\label{eq:decoupling}
\Gamma(T_{dec}) = H(T_{dec})
\ee
where
\be
\label{eq:rates}
\Gamma(T) = n_{\textbf{\scriptsize{e}}}\sigma_{T} =X_{\textbf{\scriptsize{e}}} n_{DM}\sigma_{T} \\
\label{eq:hubble}
H(T)=\frac{5}{3}\sqrt{g_*}\frac{T^2}{M_{pl}}
\simeq 5.5 \frac{T^2}{M_{pl}}
\ee
and $\sigma_{T}$ is the Thomson cross section, $\sigma_{T}=8/3 \pi
(\alpha_D / m_{\textbf{\scriptsize{e}}})^2$.  Using
(\ref{eq:decoupling}) - (\ref{eq:hubble})
the decoupling temperature is given by\footnote{Most of the allowed parameter space requires
  $B<\Lambda_{QCD}$, thus we take $g_* = 12$.}  
\be
\label{eq:dec_temp}
T_{dec} \simeq 10 \left(\frac{0.1}{\alpha_D}\right)^2 \left(\frac{0.1}{X_{\textbf{\scriptsize{e}}}} \right)
\left(\frac{m_{\textbf{\scriptsize{e}}}}{1 \,\gev}\right)^2 \left(\frac{m_{\textbf{\scriptsize{H}}}}{100 \,\gev}\right) \kev.
\ee
A full analysis of the power spectrum is left to
future work (see \cite{Feng:2009mn,Loeb:2005pm,Hooper:2007tu} for
discussions of similar effects), however, a rough estimate of the comoving wavenumber that is
damped due to the radiation interaction is
\be
\label{eq:k_Damp}
k_{damp} \lsim \frac{1}{\eta_{dec}}
\ee
where $\eta_{dec}$ is the conformal time of decoupling.  This can be
written in terms of today's temperature, $T_0$, and the decoupling temperature $T_{dec}$
\be
\label{eq:eta_Dec}
\eta_{dec} = 2t_{dec}(1+z_{dec}) \simeq \frac{M_{pl}}{T_0 T_{dec}}.
\ee
The mass scale which survives damping, and which eventually characterizes the
first dark matter clumps, can be written as  
\be
\label{eq:m_grow}
  M_{grow} & > & \frac{4 \pi}{3} \left(\frac{\pi}{k_{damp}}\right)^3
  \Omega_{DM}\rho_{crit} \nonumber \\
  & \gsim & (10^3-10^6) \left(\frac{T_{dec}}{10 \, \kev}\right)^{-3} M_{\odot}
\ee
where $\rho_{crit}$ is today's value and the largest value of
  $M_{grow}$ corresponds to $k_{dec}=\eta^{-1}_{dec}$.  The large range of $M_{grow}$ occurs because we have allowed for an order
of magnitude error in $k_{dec}$.
Thus, we find that atomic dark matter can have much less power
on small scales compared to a conventional CDM
WIMP if there is a sizeable fraction of free electrons.  In particular, weak-scale dark atoms can be consistent with the
observed populations of intermediate mass dark halos in the Milky Way \cite{Gilmore:2007fy}.
We emphasize that simulations are necessary to evaluate the
detailed power spectrum and satellite popluations predicted by atomic
dark matter. However, it is clear that even the simplest atomic
dark matter system can significantly impact structure formation.

\section{Direct Detection}
Atomic dark matter, as thus far considered, is secluded from the
standard model. While the cosmology of atomic dark matter is
interesting in its own right, it
naturally lends itself to inelastic scattering because of energy
level quantization. This offers an exciting possible explanation of
the DAMA data \cite{Bernabei:2008yi}. \\

The unperturbed energy levels of hydrogen are 
\begin{equation}
\label{eq:hydrogenE}
E_n=\frac{\alpha_D^2 \mu_{\textbf{\scriptsize{H}}}}{2\,n^{2}}.
\end{equation}
One might hope that the DAMA scale -- $\order(100\,\kev)$ -- could be generated by
energy differences between levels with different principle quantum
numbers.  Generically the rate of elastic scattering will be greater
than that of inelastic scattering.  However, predominantly inelastic
scattering could be enforced by setting $m_{\textbf{\scriptsize{p}}}=m_{\textbf{\scriptsize{e}}}=2\mu_{\textbf{\scriptsize{H}}}$.  In this
case, the first Born term for elastic scattering vanishes.  Unfortunately, efficient recombination in such a
scenario forces one to consider $m_{\textbf{\scriptsize{H}}} \sim$ \gev, which is too
small to account for the recoil energies measured by DAMA.  Nevertheless, atoms have a rich structure and the allowed parameter space for
viable recombination naturally leads to \textit{hyperfine splittings}
on the order of 100 \kev~for weak-scale hydrogen.  The hyperfine splitting is given by
\begin{equation}
\label{eq:hpf}
E_{hf} = \frac{2}{3} g_{\textbf{\scriptsize{e}}} g_{\textbf{\scriptsize{p}}} \alpha_D^4 \mu_{H} \frac{m_{\textbf{\scriptsize{e}}}}{m_{\textbf{\scriptsize{p}}}}
\end{equation}
where $g_{\textbf{\scriptsize{e}}}, g_{\textbf{\scriptsize{p}}}$ are
the gyromagnetic ratios of the dark electron and dark proton, which we take to be equal to two.  \\

Exploiting this scale requires a scattering process which is dependent
on the spins of the dark atom's constituents.  This can be
accomplished with a broken $U(1)_X$ which is axially coupled to the
dark matter and mixed with the standard model hypercharge as in \cite{Alves:2009nf} 
\be
\label{eq:kineticmixing}
\mathcal{L}_{mix} = \epsilon\,X^{\mu\nu}B_{\mu\nu}.
\ee
Having an
axial coupling in the dark sector and a vector coupling to the
standard model will ensure that the dominant scattering process
changes the dark atom spin state by one unit.
After integrating out the Z boson and diagonalizing the gauge kinetic
terms, the Lagrangian is

\begin{eqnarray}
\label{eq:fullL}
\mathcal{L} &=& \mathcal{L}_{SM}+\mathcal{L}_{DM}+\mathcal{L}_{Dark\,Gauge}\nonumber\\
\mathcal{L}_{DM} &=&\overline{\Psi}_{\textbf{\scriptsize{p}}}(i\,\fslash{\partial} 
                            - g_5 \, \gamma_5 \,\fslash{X}
			    + g\fslash{A} 
			    + m_{\textbf{\scriptsize{p}}})
                     \Psi_{\textbf{\scriptsize{p}}}
		     + 		     
		     \overline{\Psi}_{\textbf{\scriptsize{e}}}(i\,\fslash{\partial} 
                            + g_5 \,\gamma_5 \, \fslash{X}
			    -  g\fslash{A}
			    +m_{\textbf{\scriptsize{e}}})
                     \Psi_{\textbf{\scriptsize{e}}}
		     -
		     \frac{\epsilon\,s_w}{m_Z^2}\,J_{Z\,\mu}J_D^{\mu}  
		     \nonumber \\
\mathcal{L}_{Dark\,Gauge} &=& - \frac{1}{4}A_{\mu\nu}^2 
                              - \frac{1}{4}X_{\mu\nu}^2 
			      - \left(\epsilon\,c_w J^{\mu}_{EM} 
			      + \epsilon\,s_w\left(\frac{M_X}{m_Z}\right)^2 J_Z^{\mu} \right)X_{\mu} 
			      + \frac{M_X^2}{2}X^2.
\end{eqnarray}
The parameters $c_w$ and $s_w$
are the cosine and sine of the weak mixing angle.  The dark current
$J^{\mu}_D$ is 
\be
J_D^{\mu} = -g_5\,\overline{\Psi}_{\textbf{\scriptsize{p}}}\gamma^{\mu}\gamma_5\Psi_{\textbf{\scriptsize{p}}} + g_5\,\overline{\Psi}_{\textbf{\scriptsize{e}}}\gamma^{\mu}\gamma_5\Psi_{\textbf{\scriptsize{e}}},
\ee
and $J^{\mu}_{EM}$ and $J^{\mu}_{Z}$ are the standard model electromagnetic and
weak neutral currents, respectively. \\

The calculation of the direct detection scattering cross section is
organized as follows.  First, we derive the non-relativistic
interaction Hamiltonian for dark atoms and standard model nucleons
from Eq.\,(\ref{eq:fullL}).  Second, we use this to calculate the
differential cross section for a dark atom to scatter from a spin
singlet to a spin triplet state off of a standard model nucleon and append a form factor to account for recoil of the entire nucleus.  Third,
we rewrite the resulting rate in terms of the nuclear recoil $E_R$.
Finally, we convolve the recoil rate with the dark matter velocity
distribution.

\subsection{Non-relativistic Interaction Hamiltonian}

In order to calculate the scattering cross section for dark atoms off
of standard model nuclei we derive the interaction Hamiltonian by taking the non-relativistic
limit of the current-current interaction 

\be
\label{eq:currentcurrent}
\mathcal{A} = <J^{\mu}_D D_{\mu \nu} J^{\nu}_{Y}>,
\ee
where $D_{\mu \nu}$ is the Coulomb gauge propagator for $X$.  The leading behavior of Eq.\,(\ref{eq:currentcurrent}) is 

\begin{eqnarray}
\label{eq:ccnonrel}
\mathcal{A}  &\simeq& 
  \frac{ g_5\,\epsilon\,c_w\,e}{\vec{q}^{\,2} + M_X^2} \, \chi^{'\dagger}_{s'}
  \left[ 
    \frac{\vec{\sigma}_{\textbf{\scriptsize{e}}} \cdot \vec{p}_{\textbf{\scriptsize{e}}}}{m_{\textbf{\scriptsize{e}}}} +
    \frac{\vec{\sigma}_{\textbf{\scriptsize{e}}} \cdot \vec{q}}{2\,m_{\bolde}} + 
    \frac{\vec{\sigma}_{\textbf{\scriptsize{e}}} \cdot \vec{q}}{2\,m_n} - 
    \frac{\vec{\sigma}_{\textbf{\scriptsize{e}}}\cdot\vec{p}_{n}}{m_n} + 
    \frac{\vec{\sigma}_{\textbf{\scriptsize{p}}}\cdot\vec{p}_{n}}{m_n} - 
    \frac{\vec{\sigma}_{\textbf{\scriptsize{p}}}\cdot\vec{q}}{2\,m_n}\right. \nonumber\\
     &+&\left.\frac{(\vec{\sigma}_{\textbf{\scriptsize{e}}} \cdot \vec{q})(\vec{q}\cdot\vec{p}_{n})}{m_n(\vec{q}^{\,2}+M_{X}^{2})} -
    \frac{(\vec{\sigma}_{\textbf{\scriptsize{e}}} \cdot \vec{q})\vec{q}^{\,2}}{2\,m_n(\vec{q}^{\,2}+M_{X}^{2})} - 
    \frac{(\vec{\sigma}_{\boldp} \cdot \vec{q})(\vec{q}\cdot\vec{p}_{n})}{m_n(\vec{q}^{\,2}+M_{X}^{2})} +
    \frac{(\vec{\sigma}_{\boldp} \cdot \vec{q})\vec{q}^{\,2}}{2\,m_n(\vec{q}^{\,2}+M_{X}^{2})}
  \right] 
  \chi_s \, \xi^{'\dagger}_{r'} \xi_r.
\end{eqnarray}
where $m_n$ is the nucleon mass, $\vec{p}_{\bolde}$ is the initial
momentum of the dark electron, that of the nucleon is $\vec{p}_n$ and
$\vec{q}$ is the momentum conjugate to the relative coordinate between
the electron and nucleon. $\chi_s\,\textrm{and}\,\chi_{s'}$ are the
initial and final spin states of the atom and can be written in the
form: $\chi_{Atom} =
\chi_{\textbf{\scriptsize{p}}}\otimes\chi_{\textbf{\scriptsize{e}}}$.
The dark matter spin operators are $\vec{S}_{e,\,p} = \mathbbm{1}
\otimes \vec{\sigma}_{\textbf{\scriptsize{e}}}
/2,\vec{\sigma}_{\textbf{\scriptsize{p}}} /2 \otimes \mathbbm{1}$.
$\xi_r \, $ and $\xi_{r'}$ are the initial and final spin states of the standard model
nucleon. In the following analysis we consider
proton masses of $\mathcal{O}(100\,\gev)$ and electron masses of
$\mathcal{O}(1\,\gev)$, so we have ignored terms suppressed by
$m_{\textbf{\scriptsize{p}}}$.  We have also dropped terms suppressed by $M_Z$.
Finally, we have omitted terms which depend on the spin of the
standard model nucleon, as we expect these terms to contribute
\textit{incoherently} to the overall scattering cross section and
hence be suppressed by the atomic number of the nucleus. 

\subsection{Inelastic Dark Atom - Nucleus Scattering}

The cross section in the center of mass of the hydrogen-nucleus system is given by \cite{Sakurai:1995, Shankar:1994}   

\be
\label{eq:xsectdef}
\frac{d\,\sigma_{hf}}{d\,\Omega}\equiv\frac{d\sigma}{d\,\Omega}(S = 0
\rightarrow S = 1) = \frac{\mu_{NA}^2}{4
  \pi^2}\left|\frac{k'}{k}\right|\left|\bra{\textbf{p}'_{\boldH} ,
  \textbf{p}'_N, N', H'}\,\hat{H}_{int}\,\ket{\textbf{p}'_{\boldH},
  \textbf{p}_N, N, H}\right|^2. 
\ee

Here $\hat{H}_{int}$ is the interaction Hamiltonian, which is obtained
from Eq.\,(\ref{eq:ccnonrel}) by Fourier transforming $\vec{q}$ to position
space. The prefactor contains
$\mu_{NA}=(m_{\boldH}+M_N)/(m_{\boldH}+M_N)$ which is the reduced mass
of hydrogen and the nucleus, and the momenta $\vec{k}$
and $\vec{k}\,'$ which are the initial and final momenta conjugate to the relative
coordinate between the atom and the nucleus.  Our basis is the hydrogen atom 
momentum $\{\textbf{p}_{\boldH} \, , \, \textbf{p}^{'}_{\boldH}\}$,  the nucleus momentum, $\{\textbf{p}_{N}
\, , \,  \textbf{p}^{'}_{N}\}$, and the internal states of
hydrogen and the nucleus, $\{A \, , \, A'\}$ and $\{N \, , \; N'\}$. 
The explicit evulation of the matrix element in Eq.\,(\ref{eq:xsectdef}) is
complicated by the fact that there are four particles in the incoming
and outgoing states. The hydrogen atom and the 
nucleus are both free particles while the electron and nucleon are
bound the respective free particle motion. Since the scattering
centers, the electron and nucleon, do not correspond to the
coordinates of free particle motion, hydrogen's center of mass and the
nucleus's center of mass, the matrix element in Eq.\,(\ref{eq:xsectdef})
contains an atomic form-factor in addition to the usual nuclear form
factor. We find, ignoring terms suppressed by $m_{\bolde}/m_{\boldH}$
and $m_n/M_N$, the cross section to be

\begin{eqnarray}
\label{eq:xsect1}
\frac{d\,\sigma_{hf}}{d\,\Omega} &=& \left| \frac{k'}{k}
\right|\left(\frac{2\,\mu_{NA}^2}{\pi^2}\right)\left(\frac{
  g_5\,Ze\,\epsilon\,c_w} {q^2 + M_X^2}\right)^2 
  \left| \chi^{'\dagger} _{s'}\,
    \left[
      F_{el}(q^2) \frac{\vec{S}_{\textbf{\scriptsize{e}}} \cdot\vec{q}}{\mu_{ne}} -
      F_{el}(q^2) \frac{(2\,\vec{S}_{\textbf{\scriptsize{e}}} \cdot\vec{q})\,q^{2}}{m_n\,(q^{2}+ M_{X}^{2})}\right.\right. \nonumber \\  
      &-&\left.\left.\frac{\vec{S}_{\textbf{\scriptsize{p}}}\cdot\vec{q}}{m_n}
      + \frac{(2\,\vec{S}_{\boldp} \cdot\vec{q})\,q^{2}}{m_n\,(q^{2}+
	M_{X}^{2})}
    \right]
 \chi_{s} \right|^2\, 
\left|F_{\textbf{\scriptsize{H}}}(q^2)\right|^2.
\end{eqnarray}

The electron form factor $F_{el}(q^2)$ is found to be  

\be
\label{eq:atomicff}
F_{el}(q^2) &=& \bra{0} e^{i \vec{q} \cdot \vec{r}_{\textbf{\scriptsize{e}}}} \ket{0} =  \left(1
+ \frac{q^2 a_0^2}{4} \right)^{-2}
\ee
where $\left|0\rangle\right.$ is the ground state of the
dark atom.  The
function $F_{H}(q^2)$ is the Helm nuclear form factor which accounts for
the overlap between nucleon and \textit{nuclear}
states \cite{Helm:1956zz,Jungman:1995df}. We have averaged over
initial nucleon spin and summed over final nucleon spin.
Summing over the final atomic spin states and using
Eq.\,(\ref{eq:atomicff}) we have
\be
\label{eq:xsect2}
\frac{d\,\sigma_{hf}}{d\,\Omega} = \frac{M_N^2}{2}\,\left| \frac{k'}{k}\right|
\left(\frac{g_5\,Ze\,\epsilon\,c_w}{\pi}\right)^2\left(\frac{F_{el}\,F_{H}\,G(q^{2})}
  {M_X^2\left(1+\frac{q^2}{M_X^2}\right)}\right)^2\,\left(\frac{q}{\mu_{ne}}\right)^2;
  \ee
we have defined the function $G(q^{2})$ as follows

\be
G(q^{2}) \equiv 1 + \frac{\mu_{ne}}{m_n}\left[F_{el}^{-1} - (1+F_{el}^{-1}) \frac{q^{2}}{q^{2}+M_{X}^{2}}\right].
\ee

We rewrite the above in the following form

\be
\label{eq:xsectER}
\frac{d\,\sigma_{hf}}{d\,E_R} &=&
\frac{4\,Z^2\,\alpha}{\mu^2_{ne}\,f^4_{eff}}\frac{M^2_N}{v^2_{rel}}\frac{E_R\,F^2_{H}\,F_{el}^2}{(1+2M_NE_R/M_X^2)^2} G^{2}(E_{R})
\\
f_{eff}^4 &\equiv& \frac{M_X^4}{2\,(g_5\,\epsilon\,c_w)^2},
\ee
where we have defined the scale $f_{eff}$ for compactness and for comparison to Ref.\,\cite{Alves:2009nf}.

\subsection{Modulated Nuclear Recoil Rate}

The amplitude of the modulated recoil rate at a detector with $N_T$ target
nuclei of mass $M_N$ is given by  

\be
\label{eq:dRdER_def}
\left.\frac{dR}{dE_R} \equiv \frac{1}{2}\,N_T \frac{\rho}{m_{DM}}
\int_{v_{min}}^{v_{esc}}\, dv_{rel} \, v_{rel} \, f(v_{rel})
\frac{d\,\sigma_{hf}}{dE_R}\right|^{June}_{January},
\ee

\noindent where $v_{rel}$ is the relativity velocity between the dark
  atom and the nucleus, $\rho = 0.3 ~ $GeV/cm$^3$ is the local dark matter
density and we use $v_{esc} = 550~$km/s for the dark matter
escape velocity -- see Ref. \cite{Lewin:1995rx}.  The lower bound on the velocity integration is the minimum relative 
velocity that can produce a given recoil energy

\be
\label{eq:vmin}
v_{min} \equiv \sqrt{\frac{1}{2\,M_N\,E_R}}\left(\frac{M_N+M_{\textbf{\scriptsize{H}}}}{M_{\textbf{\scriptsize{H}}}}E_R + E_{hf}\right).
\ee

Following Ref's.\,\cite{Lewin:1995rx,Savage:2006qr}, the velocity distribution in Eq.\,(\ref{eq:dRdER_def}) is
approximated by

\be
\label{eq:veldist}
f(v) = \left\{
     \begin{array}{lr}
       \frac{1}{N} \left(\frac{1}{\pi v_0^2}\right)^{3/2} e^{-
       v^2/v_0^2} ,& ~ \textrm{for} ~ v < v_{esc} \\
       0 ,& ~~ \textrm{for} ~ v \geq v_{esc}.
     \end{array}
   \right.
\ee
\noindent The normalization is 
\be
N = \textrm{erf}\left(\frac{v_{esc}}{v_0}\right) -
\frac{2}{\sqrt{\pi}} \, \frac{v_{esc}}{v_0} \,
e^{-\left(v_{esc}/v_0\right)^2},
\ee with $v_0 = 220~$km/s.\\

\begin{figure}[t]
\begin{center}
\includegraphics[scale = 0.3, angle=270]{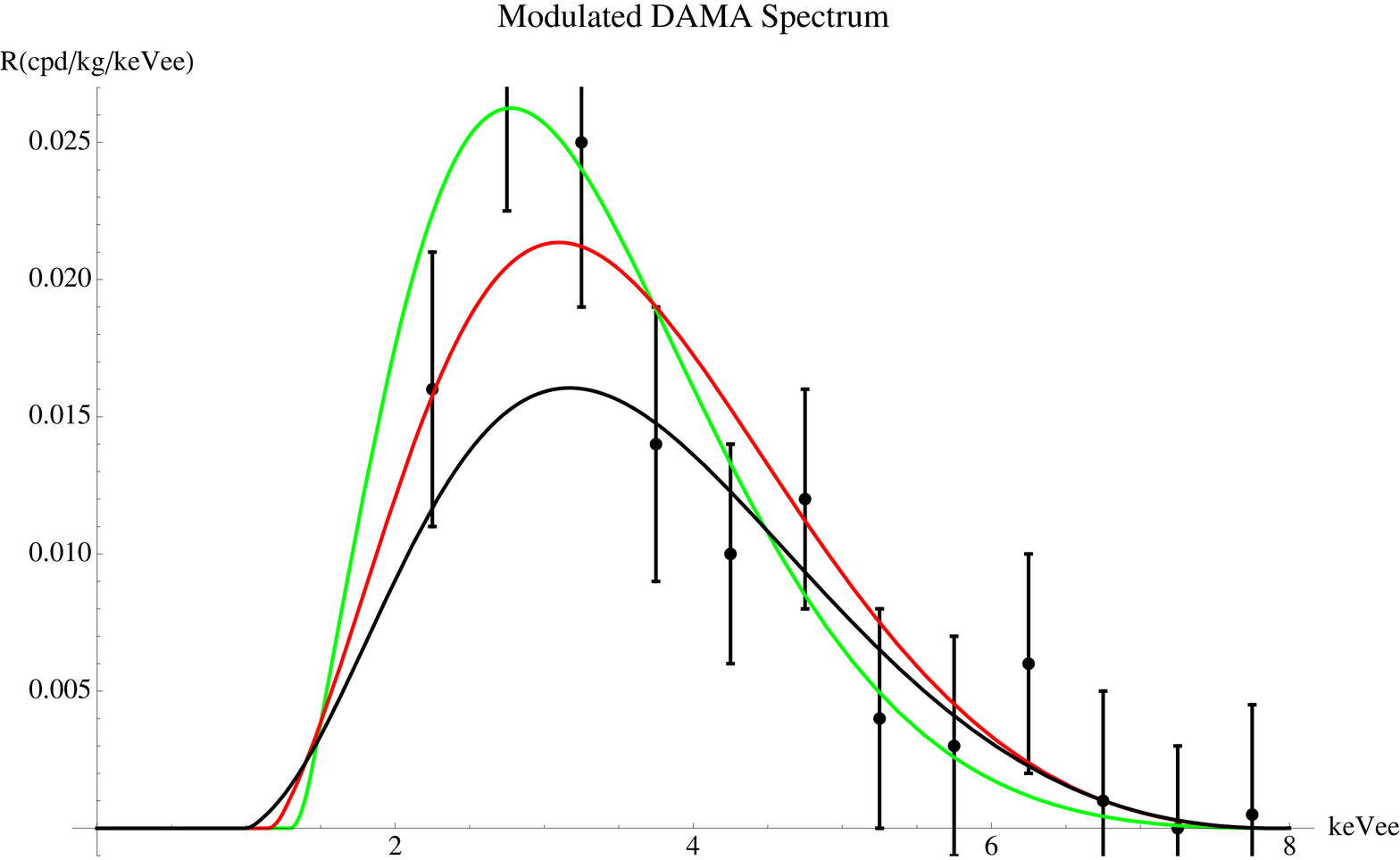}
\caption{Examples of the modulated spectrum at DAMA defined in
  Eq.\,(\ref{eq:dRdER_def}) complete with the data points from the DAMA
  and DAMA/LIBRA experiments.  The curves correspond to the following
  choice of parameters from left to right in order of their rise from
  zero (black, red and green in color order)  $m_p = 200, 100,
  70~$\gev ;~$m_e \simeq 2.1, 1.7, 1.7~$\gev ;~$f_{eff} \simeq 103, 92, 67~$\gev.  The hyperfine splittings are all about $111~$\kev. }
\label{fig:DAMAmodspec}
\end{center}
\end{figure}

Figure \ref{fig:DAMAmodspec} is an example of the modulated
count rate at DAMA as defined in Eq.\,(\ref{eq:dRdER_def}) with the
data points and reported uncertainties from DAMA and DAMA/LIBRA
\cite{Bernabei:2008yi}.  We
have plotted the modulated spectrum for three choices of the
set of parameters
$(m_{\textbf{\scriptsize{p}}},\,m_{\textbf{\scriptsize{e}}},\,\alpha_D,\,M_X,\,g_5\,\textrm{and}\,\epsilon)$ which
satisfy the rather stringent list of constraints enumerated below.
Note the linear dependence on $E_R$ and the presence of an atomic form
factor in Eq.\,(\ref{eq:xsectER}). Although the first term tends to push
the peak toward larger values of $E_R$, the atomic form factor turns
off scattering when $qa_0 \sim 1$.

\subparagraph{Mixing With The Standard Model:} 

Perhaps the harshest constraints are on the mass of the
axial $U(1)$ and its kinetic mixing with the standard model, since the
direct detection cross section is roughly proportional to
$\epsilon^2/M_X^{4}$.  The one loop contribution of $X$ to the anomolous
magnetic moment of the muon is
\be
\label{eq:oneloopmix}
a_{\mu}^{X} = \frac{\alpha\,\epsilon^{2}}{2\,\pi} \int_{0}^{1}dz\,\frac{2\,m_{\mu}^{2}\,z(1-z)^{2}}{m_{\mu}^{2}\,(1-z)^{2}+m_{X}^{2}\,z}.
\ee
As discussed in Ref.\,\cite{Pospelov:2008zw}, regardless of how one treats
the hadronic contribution to the theoretical prediction of $a_{\mu}$, the $X$ boson's
one loop contribution must satisfy
\be
a_{\mu}^{X} &\leq& 7.4\times10^{-9}.
\ee
In order to be conservative we restrict ourselves to (see Figure 1 in
Ref.\,\cite{Pospelov:2008zw})
\be
M_X &\geq& 100\,\mev \, \,
\textrm{and}\nonumber \, \, 
\epsilon^2 \lsim 10^{-5}.
\ee

Mixing between the massless gauge boson in the dark sector and the
photon is not induced by loops in our theory, and we nominally set it zero.
A constraint on this mixing, $\epsilon'$, can be derived from bounds on its
contribution to the anomolous magnetic moment of the electron. The
constraint is  $\epsilon' < \order$($10^{-4}$)
\cite{Pospelov:2008zw}.  Astrophysical constraints also exist, but are much less restrictive for the range of electron masses we are considering \cite{Davidson:2000hf}.

\subparagraph{Sufficient Recombination:}The residual ionized dark
matter will scatter \textit{elastically} as it does not cost any
energy to flip a free spin.  
Efficient recombination and a hyperfine
splitting consistent with DAMA imply that the typical electron mass
is $\mathcal{O}(1 \, \gev)$ and therefore too small to induce observable
nuclear recoils.  The strongest constraints on direct
detection of the free dark protons come from CDMS \cite{Ahmed:2008eu,Chang:2008gd}. With
a net exposure of 174.7 kg-d, the CDMS experiment allows $5.3$ signal
events at 90\% confidence level.  To be consistent with the bounds
from direct detection\footnote{The discussion in this section actually
  puts a bound on the \textit{local} ionized fraction.  We assume for
  simplicity -- here and throughout the paper -- that the distribution
  of ionized dark matter matches that of atomic dark matter.
  However, due to the presence of a long-ranged force, it may be that the
  ionized distribution is very different from the typical dark matter
  halo.  A full
  N-body simulation of a multiple species halo is beyond the scope of
  the present work, so for now we ignore this interesting possibility.}
 we demand that $X_{\textbf{\scriptsize{e}}} \leq 10^{-4}$. 

\subparagraph{Energy Level Corrections Due to the Axial U(1):}
The proton-electron interaction due to the broken axial U(1) is a
perturbation to the hydrogen Hamiltonian and gives a correction to the hyperfine level splitting.  The
correction 
 is given by  

\be
\label{eq:deltaehf}
\delta E_{hf} \sim \bra{0}\,\frac{g_5^2}{4\pi} \frac{e^{-M_X
    r}}{r}\ket{0} = \frac{g_5^2}{4\pi\,a_0}(1+a_0 M_X)^{-2}. 
\ee
Requiring $ \delta E_{hf}  \ll E_{hf}$ gives
\be
\label{eq:deltaehfapprox}
g_5^2 \ll \frac{32\,\pi}{3}\,\alpha_D^3\,\frac{m_{e}}{m_{\textbf{\scriptsize{p}}}}\left(1+ \frac{M_X}{\alpha_{D}\,\mu_{H}}\right)^{2}.
\ee

\begin{figure}[t]
\begin{center}
\subfigure{
  \includegraphics[scale=1.0]{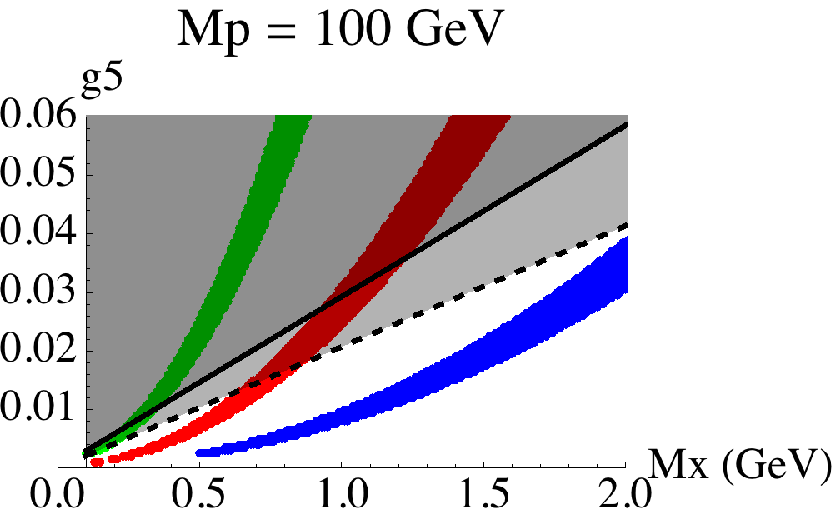}}
 \subfigure{
  \includegraphics[scale=1.0]{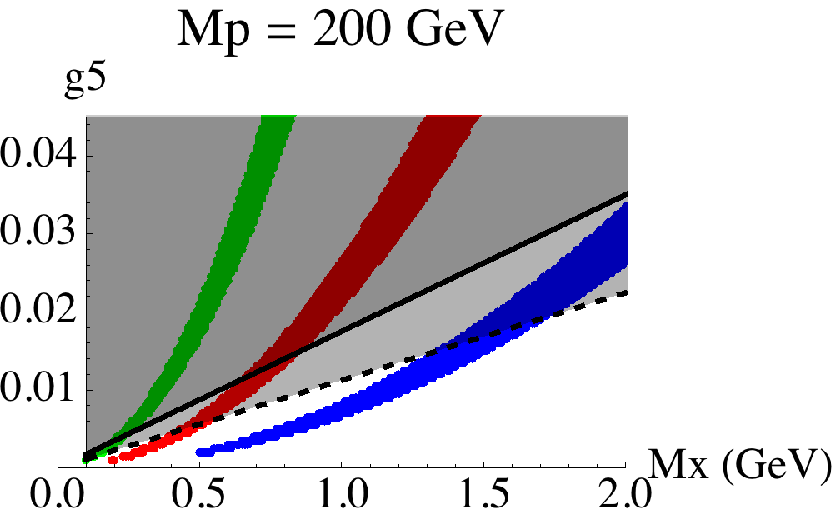}}
\caption{The allowed parameter space for $M_X$ and $g_5$ for two
  values of the dark proton mass.  The other atomic parameters --
  $\alpha_D$ and $m_{\textbf{\scriptsize{e}}}$ -- have been chosen so that $E_{hf} \simeq
  111\,\kev$ and $X_{\textbf{\scriptsize{e}}} \leq 10^{-4}$.  On each plot, the displayed
  values of $\epsilon^2$ are
  $10^{-6},\,10^{-5}\,\textrm{and}\,10^{-4}$ from left to right
  (green, red and blue in color order).  To be consistent with constraints discussed in Ref.\,\cite{Pospelov:2008zw}, $M_{X}\geq 0.5\,\gev$ for $\epsilon^{2} = 10^{-4}$.  The allowed points have an
  average rate, in the 2 to 6 keVee bins at DAMA,  between
  $0.99\times10^{-2}$ and $1.63\times10^{-2}$ cpd/kg/keVee.  The
  excluded regions correspond to choices of parameters which do not
  satisfy Eq.\,(\ref{eq:massspec}); the solid black line is for $\Lambda = 1\,\tev$ and the dashed line is for $\Lambda = 10\,\tev$.}
\label{fig:Mx_g5_allowed}
\end{center}
\end{figure}

The parameter sets shown in Figure \ref{fig:DAMAmodspec} satisfy
this constraint. 
\subparagraph{Breaking the Axial U(1):}
Masses for the dark electron, dark proton and the axial gauge boson
all violate the axial U(1) symmetry.  Perhaps the simplest way to give
mass to these particles is giving a vev to a charge $+2$ scalar
$\phi_{+2}$, as in \cite{Alves:2009nf}
\be
\label{eq:symmbreakingL}
\mathcal{L} \ni \left|D_5^{\mu} \phi_{+2} \right|^2 -
\lambda \left( \left| \phi_{+2} \right|^2 - v_{+2}^2 \right)^2 + y_{p}
(\bar{p}_R\, \phi_{+2}\, p_{L} + \bar{p}_L\, \phi_{+2}^*\, p_R) +
y_{e} (\bar{e}_R\, \phi_{+2}\, e_{L} + \bar{e}_L\, \phi_{+2}^*\,
e_R).
\ee
When the scalar is at its vev, the mass spectrum is 
\begin{eqnarray}
\label{eq:massspec}
M_X &=& g_5\,v_{+2}\nonumber\\
m_{\textbf{\scriptsize{p}}} &=& y_{\textbf{\scriptsize{p}}}\,v_{+2}\nonumber\\
m_{\textbf{\scriptsize{e}}} &=&  y_{\textbf{\scriptsize{e}}}\,v_{+2}.
\end{eqnarray}

The DAMA signal requires $m_{\boldp} > m_X$ and $g_5 \gsim
\order(10^{-2})$ thus one might worry about the perturbativity of $y_{\boldp}$.
The yukawa coupling runs
according to the following one loop renormalization group evolution \cite{Machacek:1983fi}
\be
y_{\scriptsize{\textbf{p}}}(\Lambda) = \sqrt{\frac{2\,\pi^2}{\ln(\Lambda/m_{\scriptsize{\textbf{p}}})}},
\ee
which blows up at the scale $\Lambda$. If we take $\Lambda = 1\, \tev
\, \textrm{or}\, 10 \, \tev$, our parameter space is constrained as shown in
Figure \ref{fig:Mx_g5_allowed}.  In principle, the proton could be a
composite object and the axial-symmetry breaking could occur at strong
coupling (as in QCD) and not via a weakly coupled scalar.  The proton
could also carry a charge under another gauge interaction that is
relatively strong, but breaks at a TeV, thus tempering the UV behavior
of $y_{\boldp}$.  We leave explicit models of UV completions to future work.\\

Figure \ref{fig:Mx_g5_allowed} displays the
allowed parameter space for a few choices of $M_X,\,g_5\,
\textrm{and}\, \epsilon$ with $X_{\textbf{\scriptsize{e}}} \leq
10^{-4}$ level.

\section{Discussion}

Dark matter succinctly explains a number of astrophysical and
cosmological observations that are otherwise puzzling.  Standard WIMP
dark matter can accommodate the gross features of these
observations and naturally exists in models that attempt to explain
the origin of the weak-scale.  However, the typical
WIMP seems unable to explain observed small-scale structure and
tensions between direct detection experiments.  These considerations
point to the possibility of a non-minimal dark sector, which contains
more similarities to the \textit{light sector} than is typically thought.  Atomic dark
matter -- with a non-negligible ionized fraction $X_{\bolde}$ and a
new massless gauge boson -- offers the possibility of significantly
different phenomena in the dark sector than those of standard WIMPs.\\

As discussed in Section \ref{sec:cosmo}, the residual ionized fraction
can keep dark matter in equilibrium with dark radiation long enough to
smooth halo structure on small scales.  Furthermore, atomic dark
matter may have hyperfine transitions of the right size to
offer an inelastic explanation for the DAMA data.  However, having
$X_{\bolde}$ large enough to smooth out structure is inconsistent with
the simultaneous postitive DAMA signal and null CDMS signal under the assumption that the charged halo has the same distribution as the
atomic halo.  Yet, if $X_{\bolde}$
is large enough, the distribution of the ionized fraction may be
smoother and more
spherical than standard halo models as suggested in
Ref. \cite{Wandelt:2000ad,Dave:2000ar}.  If the local ionized dark matter distribution
is very different from the atomic dark matter distribution, then parameter space
exists which can explain both small-scale structure and the DAMA
signal. On the other hand, if our simple model of atomic
dark matter is the right explanation for DAMA and the ionized
components of the halo follow the distribution of the atomic dark matter, then other direct
detection experiments should see dark protons in the near
future.  Simulations of stucture formation with charged and neutral
components could shed light on these issues. 
\\

The dynamics that lead to atomic dark matter also may have other
phenomenological implications.  For example, in parts of parameter space where the ionized fraction is
large enough, $\textbf{H}_2$ molecules may form through processes
catalyzed by the residual ions, as in the SM \cite{Hirata:2006bt}
\be
\textbf{H} + \textbf{e} &\leftrightarrow& \textbf{H}^{-} +
\textbf{$\gamma$} \nonumber\\
\textbf{H}^{-} + \textbf{H} &\leftrightarrow& \textbf{H}_{2} + \textbf{e}
\ee
and
\be
\textbf{H} + \textbf{p} &\leftrightarrow& \textbf{H}^{+}_{2} + \textbf{$\gamma$}\nonumber\\
\textbf{H}^{+}_{2} + \textbf{H} &\leftrightarrow& \textbf{H}_{2} + \textbf{p}.
\ee
The existence of molecular states in the dark sector offers the
possibility of cooling mechanisms which, in the SM, are thought to be
very important for the formation of the first stars
\cite{Yoshida:2006bz}.  This raises the interesting question of
whether and to what extent compact objects, \emph{e.g.} dark stars, could form
for weak-scale dark atoms.  Moreover, if the dark photon mixes with the SM
photon, it may result in dark atomic line emissions in cosmic gamma rays.\\

We have presented a somewhat generic model of atomic dark matter.  Explicit models which explain the asymmetry abundance and which serve as ultraviolet completions of the model could potentially relate astrophysical phenomena to physics to be probed by the Large Hadron Collider.  The part of parameter space in which the measured DAMA signal is post-dicted requires the dark proton to be strongly coupled, or nearly so, at a TeV.  If strongly coupled, one could imagine additional features of the dark sector -- {\it i.e.}, a composite atomic nucleus -- which more strongly mimic our visible world.\\

 
The authors would like to thank Julian Krolik, Kirill Melnikov, Colin
Norman, and Alex Szalay for helpful discussions.  This work is
supported in part by the National Science Foundation under grant
NSF-PHY-0401513, the Department of Energy's OJI program under grant
DE-FG02-03ER4127, and the OWC.\\


\begin{thebibliography}{60}

\bibitem{Komatsu:2008hk}
  E.~Komatsu {\it et al.}  [WMAP Collaboration],
  Astrophys.\ J.\ Suppl.\  {\bf 180}, 330 (2009)
  [arXiv:0803.0547 [astro-ph]].

\bibitem{Abazajian:2004tn}
  K.~Abazajian {\it et al.}  [SDSS Collaboration],
  Astrophys.\ J.\  {\bf 625}, 613 (2005)
  [arXiv:astro-ph/0408003].

\bibitem{Navarro:1995iw}
  J.~F.~Navarro, C.~S.~Frenk and S.~D.~M.~White,
  Astrophys.\ J.\  {\bf 462}, 563 (1996)
  [arXiv:astro-ph/9508025].

\bibitem{Gilmore:2007fy}
  G.~Gilmore, M.~I.~Wilkinson, R.~F.~G.~Wyse, J.~T.~Kleyna, A.~Koch, N.~W.~Evans and E.~K.~Grebel,
  Astrophys.\ J.\  {\bf 663}, 948 (2007)
  [arXiv:astro-ph/0703308].


\bibitem{Bernabei:2008yi}
  R.~Bernabei {\it et al.}  [DAMA Collaboration],
  Eur.\ Phys.\ J.\  C {\bf 56}, 333 (2008)
  [arXiv:0804.2741 [astro-ph]].


\bibitem{Ahmed:2008eu}
  Z.~Ahmed {\it et al.}  [CDMS Collaboration],
  Phys.\ Rev.\ Lett.\  {\bf 102}, 011301 (2009)
  [arXiv:0802.3530 [astro-ph]].

\bibitem{Angle:2007uj}
  J.~Angle {\it et al.}  [XENON Collaboration],
  Phys.\ Rev.\ Lett.\  {\bf 100}, 021303 (2008)
  [arXiv:0706.0039 [astro-ph]].
  
\bibitem{Adriani:2008zr}
  O.~Adriani {\it et al.}  [PAMELA Collaboration],
  Nature {\bf 458}, 607 (2009)
  [arXiv:0810.4995 [astro-ph]].

\bibitem{:2008zzr}
  J.~Chang {\it et al.},
  Nature {\bf 456}, 362 (2008).
  
\bibitem{Torii:2008xu}
  S.~Torii {\it et al.}  [PPB-BETS Collaboration],
  arXiv:0809.0760 [astro-ph].
  
\bibitem{Abdo:2009zk}
  A.~A.~Abdo {\it et al.}  [The Fermi LAT Collaboration],
  Phys.\ Rev.\ Lett.\  {\bf 102}, 181101 (2009)
  [arXiv:0905.0025 [astro-ph.HE]].

\bibitem{Amsler:2008zzb}
  C.~Amsler {\it et al.}  [Particle Data Group],
  Phys.\ Lett.\  B {\bf 667}, 1 (2008).

\bibitem{Kaplan:2009ag}
  D.~E.~Kaplan, M.~A.~Luty and K.~M.~Zurek,
  arXiv:0901.4117 [hep-ph].

\bibitem{Farrar:2005zd}
  G.~R.~Farrar and G.~Zaharijas,
  Phys.\ Rev.\ Lett.\  {\bf 96}, 041302 (2006)
  [arXiv:hep-ph/0510079].



\bibitem{TuckerSmith:2001hy}
  D.~Tucker-Smith and N.~Weiner,
  Phys.\ Rev.\  D {\bf 64}, 043502 (2001)
  [arXiv:hep-ph/0101138].



\bibitem{Feng:2009mn}
  J.~L.~Feng, M.~Kaplinghat, H.~Tu and H.~B.~Yu,
  arXiv:0905.3039 [hep-ph].
  
  
\bibitem{Ackerman:2008gi}
  L.~Ackerman, M.~R.~Buckley, S.~M.~Carroll and M.~Kamionkowski,
  Phys.\ Rev.\  D {\bf 79}, 023519 (2009)
  [arXiv:0810.5126 [hep-ph]].


\bibitem{Alves:2009nf}
  D.~S.~M.~Alves, S.~R.~Behbahani, P.~Schuster and J.~G.~Wacker,
  arXiv:0903.3945 [hep-ph].

\bibitem{Berezhiani:2005ek}
  Z.~Berezhiani,
  arXiv:hep-ph/0508233.


\bibitem{Mohapatra:2001sx}
  R.~N.~Mohapatra, S.~Nussinov and V.~L.~Teplitz,
  Phys.\ Rev.\  D {\bf 66}, 063002 (2002)
  [arXiv:hep-ph/0111381].


\bibitem{Klypin:1999uc}
  A.~A.~Klypin, A.~V.~Kravtsov, O.~Valenzuela and F.~Prada,
  Astrophys.\ J.\  {\bf 522}, 82 (1999)
  [arXiv:astro-ph/9901240].

\bibitem{Moore:1999nt}
  B.~Moore, S.~Ghigna, F.~Governato, G.~Lake, T.~R.~Quinn, J.~Stadel and P.~Tozzi,
  Astrophys.\ J.\  {\bf 524}, L19 (1999).




\bibitem{Peebles:1968ja}
  P.~J.~E.~Peebles,
  Astrophys.\ J.\  {\bf 153}, 1 (1968).
  
\bibitem{Dodelson:2003}
  S. Dodelson, {\it Modern Cosmology} (Academic Press, San Diego, 2003).

\bibitem{Ma:1995ey}
  C.~P.~Ma and E.~Bertschinger,
  Astrophys.\ J.\  {\bf 455}, 7 (1995)
  [arXiv:astro-ph/9506072].

\bibitem{Spitzer:1978}
L.~Spitzer 1978, 
{\it Physical Processes in the Interstellar Medium}, (
Wiley, New York)


\bibitem{Spitzer:1951}
L.~J. ~Spitzer and J.~L.~Greenstein
    Astrophys.\ J.\  {\bf 114}, 407 (1952).

\bibitem{Landau:1958}
L. D. Landau and E. M. Lifshitz, {\it Quantum Mechanics: Non-Relativistic Theory} (Pergamon Press Ltd., London -- Paris, 1958.)

\bibitem{Wu:1962}
T. Wu and T. Ohmura, {\it Quantum Theory of Scattering}(Prentice-Hall Inc., New Jersey, 1962.)

\bibitem{Bransden:1983}
B.H. Bransden and C.J. Joachain, {\it Physics of Atoms and Molecules}(Addison Wesley Longman Ltd., Essex, 1983.)


\bibitem{Schultz:2008apj}
D.~R.~Schultz, P.~S.~Krstic, T.~G.~Lee and L.~C.~Raymond
    Astrophys.\ J.\  {\bf 114}, 407 (1952).

\bibitem{Krstic:1999jpb}
P.~S.~Krstic, and D~.R.~Schultz   
J.\ Phys.\ B {\bf 32} 3485 (1999)

\bibitem{Kristic:2004pra}
P.~S.~Krstic, J.~H.~Macek, S.~Yu.~Ovchinnikov, and D.~R.~Schultz
Phys.\ Rev.\ A {\bf 70} 042711 (2004)


\bibitem{Markevitch:2003at}
  M.~Markevitch {\it et al.},
  Astrophys.\ J.\  {\bf 606}, 819 (2004)
  [arXiv:astro-ph/0309303].
  
\bibitem{Randall:2007ph}
  S.~W.~Randall, M.~Markevitch, D.~Clowe, A.~H.~Gonzalez and M.~Bradac,
  arXiv:0704.0261 [astro-ph].

\bibitem{MiraldaEscude:2000qt}
  J.~Miralda-Escude,
  arXiv:astro-ph/0002050.

\bibitem{Loeb:2005pm}
  A.~Loeb and M.~Zaldarriaga,
  Phys.\ Rev.\  D {\bf 71}, 103520 (2005)
  [arXiv:astro-ph/0504112].

\bibitem{Hooper:2007tu}
  D.~Hooper, M.~Kaplinghat, L.~E.~Strigari and K.~M.~Zurek,
  Phys.\ Rev.\  D {\bf 76}, 103515 (2007)
  [arXiv:0704.2558 [astro-ph]].


\bibitem{Sakurai:1995}
J. J. Sakurai, {\it Modern Quantum Mechanics} (Addison-Wesley Publishing Company, 1995), Revised
\bibitem{Shankar:1994}
R. Shankar, {\it Principles of Quantum Mechanics} (Springer, 1994)



\bibitem{Helm:1956zz}
  R.~H.~Helm,
  Phys.\ Rev.\  {\bf 104}, 1466 (1956).


\bibitem{Jungman:1995df}
  G.~Jungman, M.~Kamionkowski and K.~Griest,
  Phys.\ Rept.\  {\bf 267}, 195 (1996)
  [arXiv:hep-ph/9506380].


\bibitem{Lewin:1995rx}
  J.~D.~Lewin and P.~F.~Smith,
  Astropart.\ Phys.\  {\bf 6}, 87 (1996).


\bibitem{Savage:2006qr}
  C.~Savage, K.~Freese and P.~Gondolo,
  Phys.\ Rev.\  D {\bf 74}, 043531 (2006)
  [arXiv:astro-ph/0607121].
  

\bibitem{Pospelov:2008zw}
  M.~Pospelov,
  arXiv:0811.1030 [hep-ph].

\bibitem{Davidson:2000hf}
  S.~Davidson, S.~Hannestad and G.~Raffelt,
  JHEP {\bf 0005}, 003 (2000)
  [arXiv:hep-ph/0001179].

\bibitem{Chang:2008gd}
  S.~Chang, G.~D.~Kribs, D.~Tucker-Smith and N.~Weiner,
  Phys.\ Rev.\  D {\bf 79}, 043513 (2009)
  [arXiv:0807.2250 [hep-ph]].
  
  
\bibitem{Machacek:1983fi}
  M.~E.~Machacek and M.~T.~Vaughn,
  Nucl.\ Phys.\  B {\bf 236}, 221 (1984).

\bibitem{Wandelt:2000ad}
  B.~D.~Wandelt, R.~Dave, G.~R.~Farrar, P.~C.~McGuire, D.~N.~Spergel and P.~J.~Steinhardt,
  arXiv:astro-ph/0006344.

\bibitem{Dave:2000ar}
  R.~Dave, D.~N.~Spergel, P.~J.~Steinhardt and B.~D.~Wandelt,
  Astrophys.\ J.\  {\bf 547}, 574 (2001)
  [arXiv:astro-ph/0006218].

\bibitem{Hirata:2006bt}
  C.~M.~Hirata and N.~Padmanabhan,
  Mon.\ Not.\ Roy.\ Astron.\ Soc.\  {\bf 372}, 1175 (2006)
  [arXiv:astro-ph/0606437].

\bibitem{Yoshida:2006bz}
  N.~Yoshida, K.~Omukai, L.~Hernquist and T.~Abel,
  Astrophys.\ J.\  {\bf 652}, 6 (2006)
  [arXiv:astro-ph/0606106].




\end{thebibliography}
\end{document}